\documentclass[preprint,iop,10pt,twocolumn] {revtex4-1}
\usepackage{graphicx}
\usepackage{dcolumn}
\usepackage{amsmath}
\usepackage{bm}
\usepackage{natbib}
\usepackage[mathlines]{lineno}%
\begin{document}
\title [Novel innovation of....] {Novel method of developing broad band AC biasing power amplifier for online turbulent feedback experiment in STOR-M tokamak}
\author{Debjyoti Basu$^{a,b}$, Masaru Nakajima$^{a}$, A.V. Melnikov$^{c,d}$, David McColl$^{a}$, Chijin Xiao$^{a}$, Akira Hirose$^{a}$}
\address{$^{a}$Plasma Physics Laboratory, University of Saskatchewan, Saskatoon, Canada}
\address{$^{b}$ Present address: Institute for Plasma Research, Bhat, Gandhinagar-382428, India}
\email{debjyotibasu.basu@gmail.com}
\address{$^{c}$ NRC Kurchatov Institute, 123182, Moscow, Russia}
\address{$^{d}$ National Research Nuclear University MEPhI, 115409, Moscow, Russia}


\date{\today}

\begin{abstract}
A pulsed oscillating power amplifier has been developed for high frequency biasing\cite{kn:deb1} and real time turbulent feedback experiment in STOR-M tokamak. It is capable to provide output peak to peak oscillating voltage of around $\pm60$V and current around 30A within frequency band 1kHz-50kHz without any distortion of any waveform signal. Overall output power is amplified by two stages power mosfet op-amp as well as nine identical push-pull amplifiers which are parallel connected in final stages. The power amplifier input signal, collected from plasma floating potential during plasma shot, is optically isolated with tokamak vessel for real time feedback experiment. Here, filtered floating potential fluctuations having band width between 5kHz-40kHz has been amplified and fed to an electrode inserted into the plasma edge to study response of plasma turbulence. It is observed that magnetic fluctuations are suppressed due to real time feedback of floating potential.
\end{abstract}

\pacs{Valid PACS appear here}
\maketitle
\section{Introduction}
Real time feedback of turbulence at the edge plasma of tokamak is promising method to control electrostatic or electromagnetic turbulence. The necessary condition to achieve burning plasmas in fusion reactor is to control anomalous transport of heat and particle where the turbulence is key factor \cite{kn:Brave} for these anomaly. Earlier experiments showed that plasma fluctuations or instabilities had been controlled and stabilized by feedback such as ion-cyclotron mode in mirror machine \cite{kn:Arsenin}, drift instabilities in linear machines \cite{kn:Parker,kn:Keen}. Also, first time MHD mode had been tried to stabilize through feed back method in ATC machine\cite{kn:Bol}. In these all experiments, feedback method dealt with a single mode or few small numbers of well defined modes in frequency domain $(\delta\omega\ll \omega)$ which are quite simpler comparing to tokamak cases.
\par The turbulence scenario in tokamak is complex comparing to linear or mirror machines where it is generated in tokamak through strong nonlinear coupling between different modes and finally create a broad band spectra with small wavelengths and spatial correlation length in perpendicular directions\cite{kn:Liewer1, kn:Callen}. The real time feedback method for electrostatic turbulence suppression has been tested in few tokamaks like TEXT\cite{kn:UCKAN}, KT-5C \cite{kn:Kan} and recently a feed back experiment had been performed in HBT-EP tokamak\cite{kn:Brooks} to stabilize MHD mode rotation. All those simulation experiments through biased electrode reveal that real time feed back of elctrostatic or magnetic may be a powerful means of active control of turbulence if proper technology of this method can be developed for reactor like machines.
To become highly fascinated of this promising technique and achieve more clear physics understanding, it is planned to perform real time feedback experiment in STOR-M tokamak starting from fundamental level. It is clear from previous experiments in TEXT tokamak \cite{kn:UCKAN} that success of active feed back control depends up on two basic requirements which are: a) gain of the feedback system, b) efficient phase shifting of broadband feedback signal from original broadband turbulent spectra. A simple electrostatic turbulent feedback control system has been developed to execute active feed back control at real time in STOR-M tokamak. This system mainly consists of a broadband power amplifier (1kHz-50kHz), optical isolator and filter. Here, feedback signal collecting from Langmuir probe has been applied to plasma through an electrode either in phase or antiphase where electrode and probe are located in same magnetic surface. The main challenging task was to develop a broadband power amplifier which can transfer wide band turbulent spectra without any distortion from its input to output.
\par In this article, elaborate discussions of experimental set up, power amplifier development, experimental outcomes and its explanations will be presented in following sections.
\section{Experimental set up}
 STOR-M is a limiter based small tokamak with circular plasma cross-section having major and minor radii $46 cm$ and $12 cm$, respectively. During real time feedback experiment floating potential signal within electronically selected time window, collected from a radial Langmuir probe in floating mode, has been amplified and fed into plasma through a rectangular electrode, made of a stainless steel plate. In this experiment, both electrode and floating potential probe which were inserted from radial port were kept in same magnetic surface$(r=0.88)$ although they are toroidally separated by $180^{0}$. It was diagnosed with floating potential probe systems, inserted from top and bottom ports and with a set of twelve Mirnov coils which ware separated toroidally by $90^{0}$ in the counterclockwise direction (top view) with respect to the location of floating potential probe system. Schematic block diagram or flow chart of an ideal active feed back system is shown in figure 1.
 \begin{figure}[h]
\center
\includegraphics[width=260pt,height=150pt]{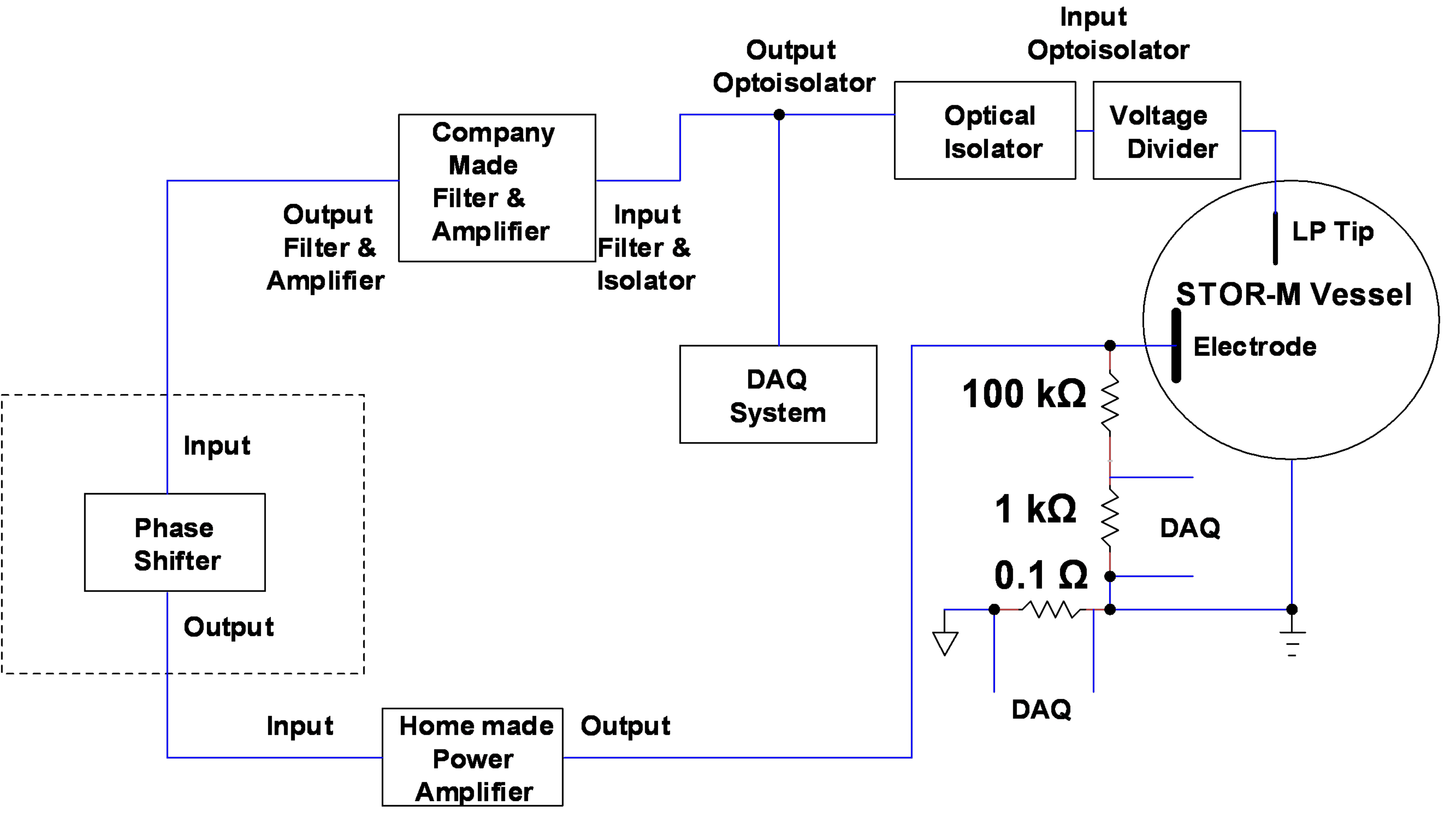}
\caption{Block diagram for ideal active feed back system.} \label{fig:1}
\end{figure}
 Block diagram clearly illustrates the step of feedback process from collecting turbulent spectra of plasma, processing of the collected spectra through feedback hardware and resend to plasma after amplification by power amplifier. The main part of ideal feedback hardware consists of a Langmuir probe in floating mode with voltage divider, optical isolator, band pass analogue filter, phase shifter, power amplifier and an electrode. In our real experiment, processed feedback signal has been sent to plasma either in phase or antiphase with collected turbulent spectra. Here, collected broadband turbulent spectra has been filtered in between 5kHz-40kHz and optical isolator is used to avoid any kind of ground loop. The power amplifier is key section of the feedback hardware and its block diagram is shown in figure 2.
\begin{figure}[h]
\center
\includegraphics[width=260pt,height=160pt]{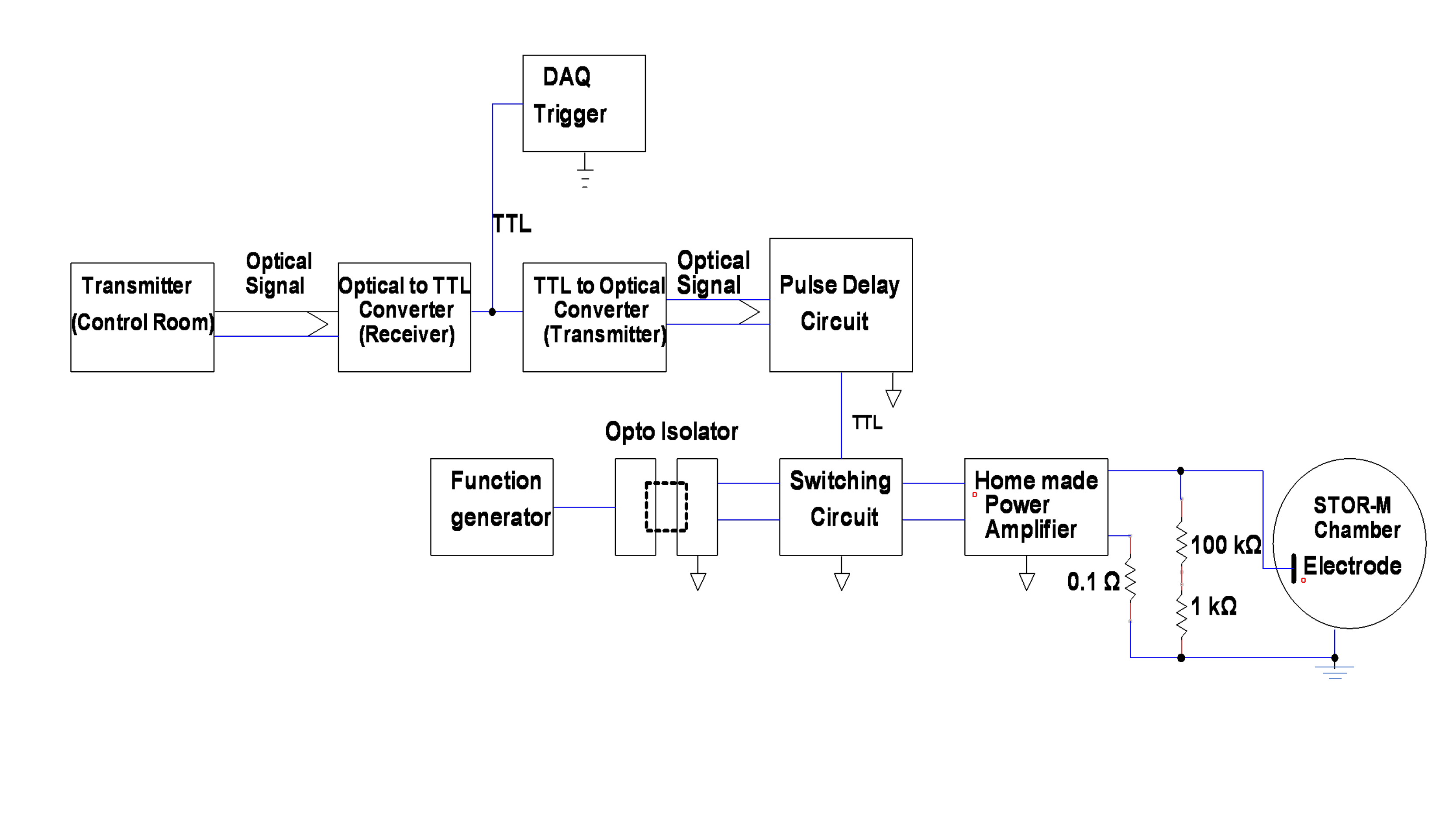}
\caption{Block diagram of broad band power amplifer.} \label{fig:2}
\end{figure}
This power amplifier has been developed for single mode operation for high frequency biasing\cite{kn:deb1} experiment as well as active turbulent feedback experiment. Its input signal has been applied through function generator for single mode operation where function generator controls waveform, amplitude as well as frequency of input applied signal. In active feed back experiment, input signal of power amplifier is filtered floating potential collected from plasma. Onset time of input signal and its active time window during plasma discharge has been controlled through a master TTL pulse with a pulse controller circuit where master TTL pulse was generated from STOR-M control room. This master TTL pulse is the input trigger pulse of pulse controller circuit which controls both of onset time as well as active time duration of applied input signal of power amplifier through an electronic solid state switch LF13202. Figure 3 shows pulse delay control circuit, optical isolator circuit as well as electronic switch circuit. These three circuits have been developed to modify conventional/recomended circuits according to experimental needs.
\begin{figure}[h]
\center
\includegraphics[width=260pt,height=180pt]{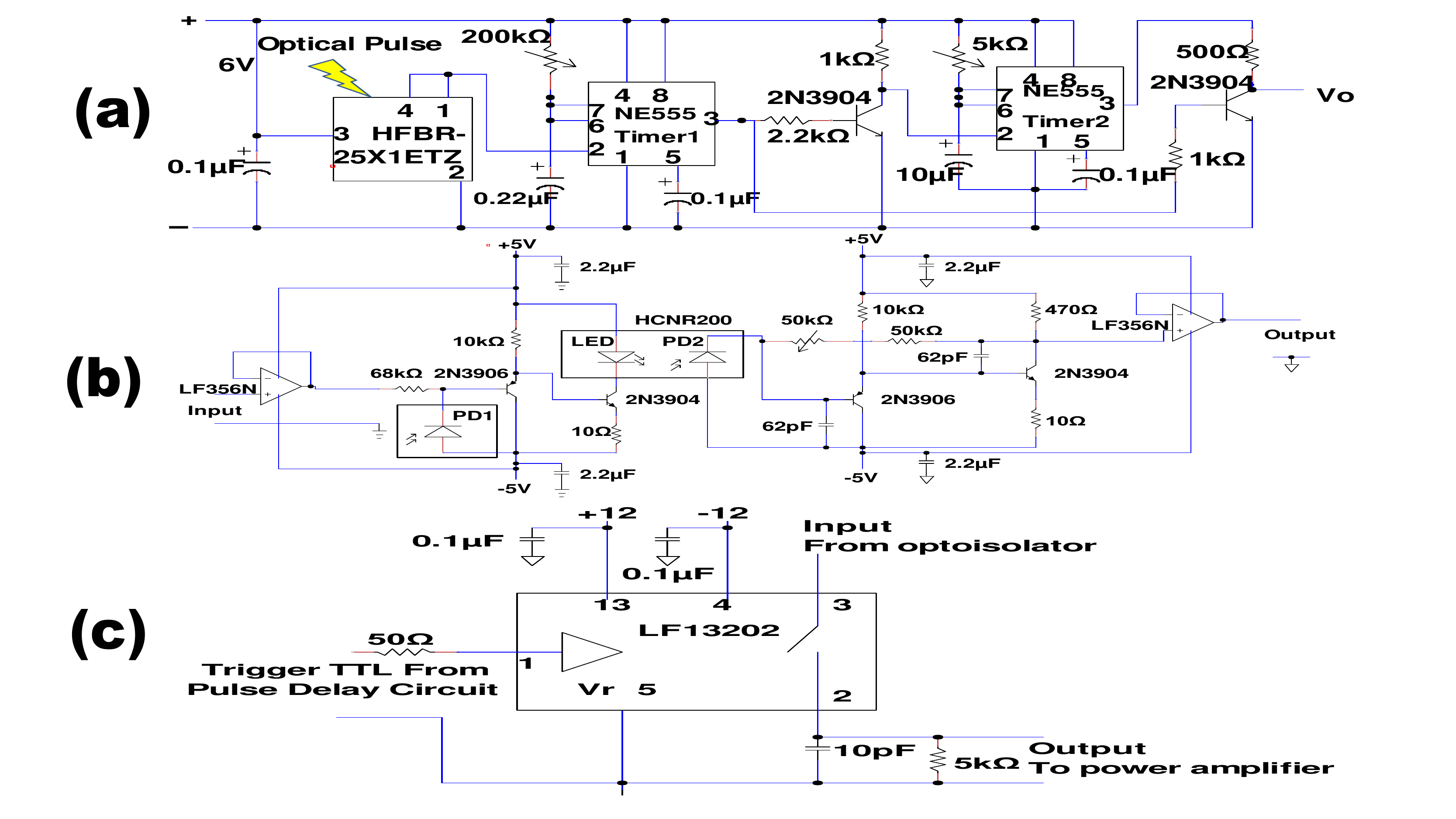}
\caption{Electronic circuit diagram of (a)pulse controller, (b)opto-isolator, (c)Solid state electronic switch.} \label{fig:3}
\end{figure}
The key part of pulse controller circuit consists of HFBR-25X1ETZ receiver, two NE555 timer and two 2N3904 NPN transistors. Here, HFBR-25X1ETZ receiver output is kept high level when there is no optical signal. When it receive optical signal by its input photodiode then its output goes to low state. Normally, both inputs of NE555 timer are kept at high level and naturally its output become low level since both are configured for monostable operation. First timer input is connected to the output of optical receiver and its output is connected bases of both transistors in this circuit. //Second timer input is connected to the collector of first transistor such that it remain high normally. Both timers are triggered at falling edge of input from higher level to lower level and output pulse width is controlled by a resister and a capacitor in series combination where pulse width is $t_{w}=1.1RC$. Finally, output pulse of pulse controller circuit is collected from collector of second transistor. Its output become low state when second stage transistor goes to turn on by both timers' high state of output. Its output provide high state when base of second transistor goes to lower level(first timer output) but its collector remain higher state(second timer output) which drive transistor to turn off. It means that transistor's output turns from lower state to higher state when first timer's output goes higher to lower state and second timer's output remains in higher state. As a result, output pulse width of pulse controller is determined by the subtracted value of second \& first timer output pulse width. Its Pulse delay and pulse width can be varied from 0.5ms to 44ms and 1ms to 49ms. It controls the activated time window of solid state switch LF13202.
\par Figure 3(b) shows optoisolator circuit to avoid any kind of ground loop formation. It is made by using of highly linear analogue optocoupler HCNR200 which has high speed with wide bandwidth from DC to 1.5MHz, highly precision with typical value 10kHz as well as maximum working insulation voltage $V_{rms}=1$kV. Here, optoisolator circuit is made on the principle of auto current feedback using HCNR200 where LED light controls photocurrent of input and output photodiodes which were used in reverse biased mode at linear region. The recommended circuit of its manual has been modified and convert into bipolar configurations. Two op-amp based voltage follower have been used to avoid any kind of signal distortion due to impedance mismatching of its input and output stages. All specifications of components have been chosen and used in the circuit according to requirements. Figure 3(c) shows that switching circuit where the electronic switch activation time is controlled by the output pulse of pulse controller circuit. The input of the switch is connected with output of optoisolator and its output is connected to the input of power amplifier. Optoisolator input signal is the signal which is coming from either plasma turbulence for active feedback experiment or function generator for single mode operation in high frequency biasing experiment. Power amplifier amplifies it and send to plasma through a biased electrode.
\par Till it is discussed about necessary peripheral controller circuits which are important for performing synchronized operation according to experimental requirements. Now, exploration about development of power amplifier will be discussed in detail in following sections.
\section{Power Amplifier}
The broadband power amplifier has been developed from basic ideas which is cost-effective. It has been made in such a way that it can easily amplify oscillating signals within frequency range 1kHz-50kHz from low power input to high power output. Here, application of a time-varying signal having voltage $\pm0.4$V and current $\pm0.5$A to the input of power amplifier will produce output signal with voltage around $\pm60$V and current around $\pm30$A without any distortion. The frequency band width is chosen in between 1kHz-50kHz because density driven drift mode calculated from edge density gradient and edge temperature of STOR-M is around 10kHz \cite{kn:deb2}as well as MHD mode is around 20kHz\cite{kn:Xiao1}. The whole power amplifier unit has two consecutive sections which amplify voltage and current respectively. The speed, bandwidth as well as slew rate of all chips used in circuit have been chosen under proper calculations such that they can serve the purpose of bandwidth as well as phase of power amplifier. Voltage and current amplification parts have been discussed in detail in following subsections.
\subsection{Voltage Amplification Unit}
Voltage amplification unit receive signal through a capacitor to remove DC offset. It has an op-amp based voltage follower and two op-amp based voltage amplifier. Each section of this unit is connected with next one through a capacitor to avoid DC offset arising from DC biasing of each op-amp. This is shown in figure 4(figure 1\cite{kn:deb1}).
\begin{figure}[h]
\center
\includegraphics[width=260pt,height=180pt]{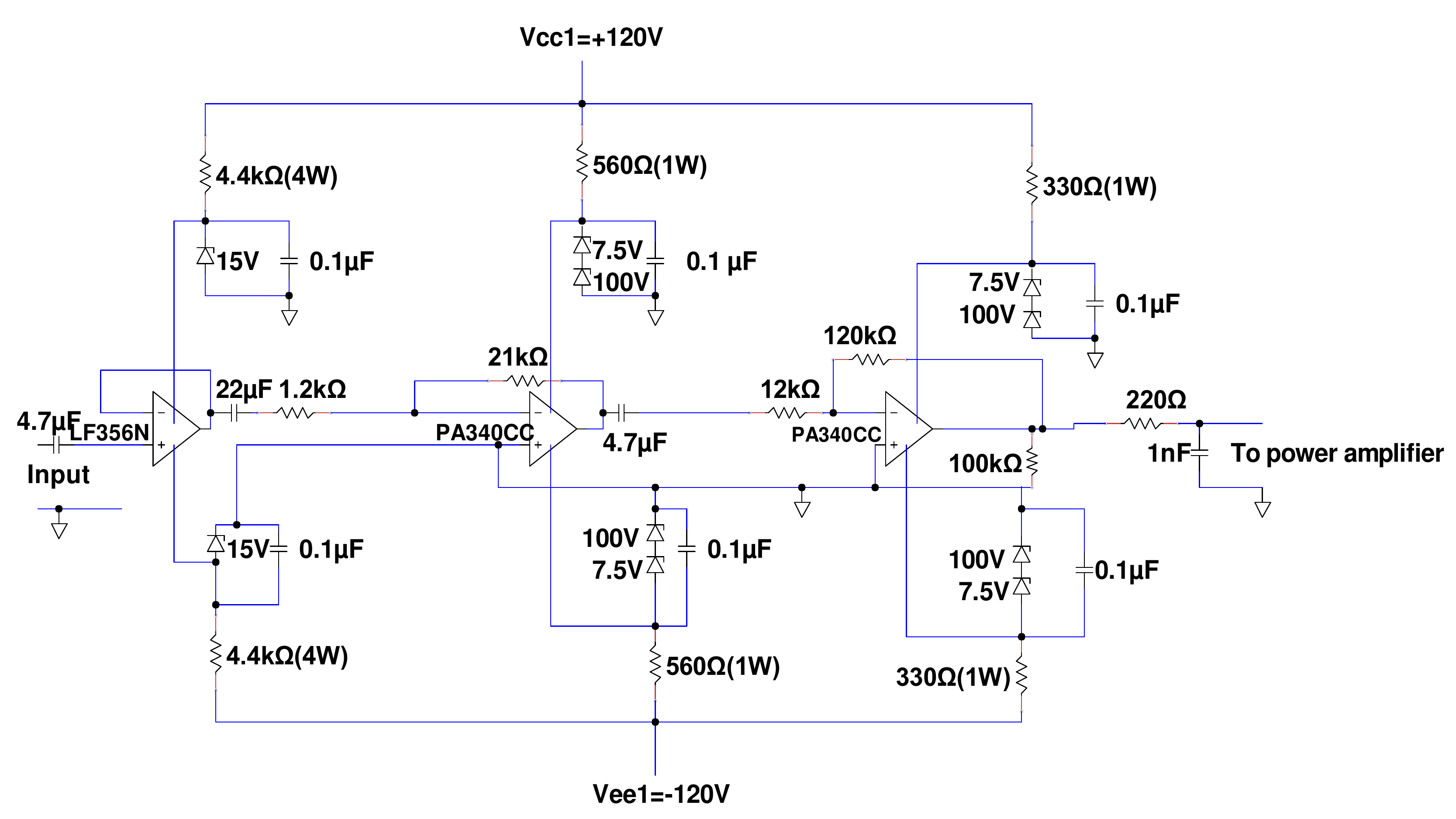}
\caption{Voltage amplification unit of power supply.} \label{fig:4}
\end{figure}
Voltage follower of this unit has been made by normal op-amp LF356N to avoid the effect of impedance mismatch with previous section. Input voltage has been amplified through high speed power mosfet op-amp PA340CC having wide frequency bandwidth and high frequency response. It has peak to peak biased voltage and peak current rating 350V \& 120mA respectively but peak to peak biased voltage and continuous current rating for safe operating area are 250V \& 60mA respectively. Its differential input voltage can be varied in between -16V to +16V. Gain Bandwidth Product(GBWP at 1MHz), slew rate(at compensation capacitance 4.7pF) and power bandwidth(at peak to peak voltage 280V) are 10MHz, 32V/$\mu$S \& 35kHz respectively. Here, amplification has been done by two stages inverting op-amp amplifiers where amplification factor of first stage and second stage are 17.5 \&  10 respectively. As a result equivalent amplification factor is around 175. Output current of voltage amplifier is limited by $330\Omega$ resistance, connected at second stage of op-amp when its biasing voltage $\pm120$V and limiting current is 38mA. Therefore, its power dissipation is 9W where continuous safe operating power dissipation is 14W.

\par To understand frequency response of this section, it needs to calculate two important parameters which are GBWP and power bandwidth. In op-amp GBWP is defined by $GBWP=GAIN\times f_{BW}$, where $f_{BW}$ is frequency bandwidth. GBWP is constant quantity for wide range of frequency for op-amp even beyond of its cutoff frequency of bandwidth. In the case of PA340CC, GBWP=10MHz when $f_{BW}=1$MHz and so, it can be said that GAIN=10 for frequency band width 1MHz. For first stage of inverting amplifier GAIN=17.5 and since GBWP=10MHz, it can be said that $f_{BW}=570$kHz. It can only predicts that voltage amplifier can amplify up to signal with frequency 570kHz. But it does not predict whether output signal will be distorted or not. To understand the frequency range of undistorted signal in output of voltage amplifier, power bandwidth(PB)is required to calculate where $PB=\frac{Slew Rate}{2\pi V_{p}}$. Here, $V_{p}$ is peak output voltage of oscillatory signal. So, calculated value of PB=72kHz where, slew rate=32V/$\mu$S \& $V_{p}=70$V. Therefor, the developed voltage amplifier able to generate undistorted output signal up to 70kHz with peak oscillating voltage 70V.
\subsection{Current Amplification Unit}
Configuration of push-pull amplifier(type `AB') has been used for power amplification section which is shown in figure 5(figure 2\cite{kn:deb1}).
\begin{figure}[h]
\center
\includegraphics[width=260pt,height=180pt]{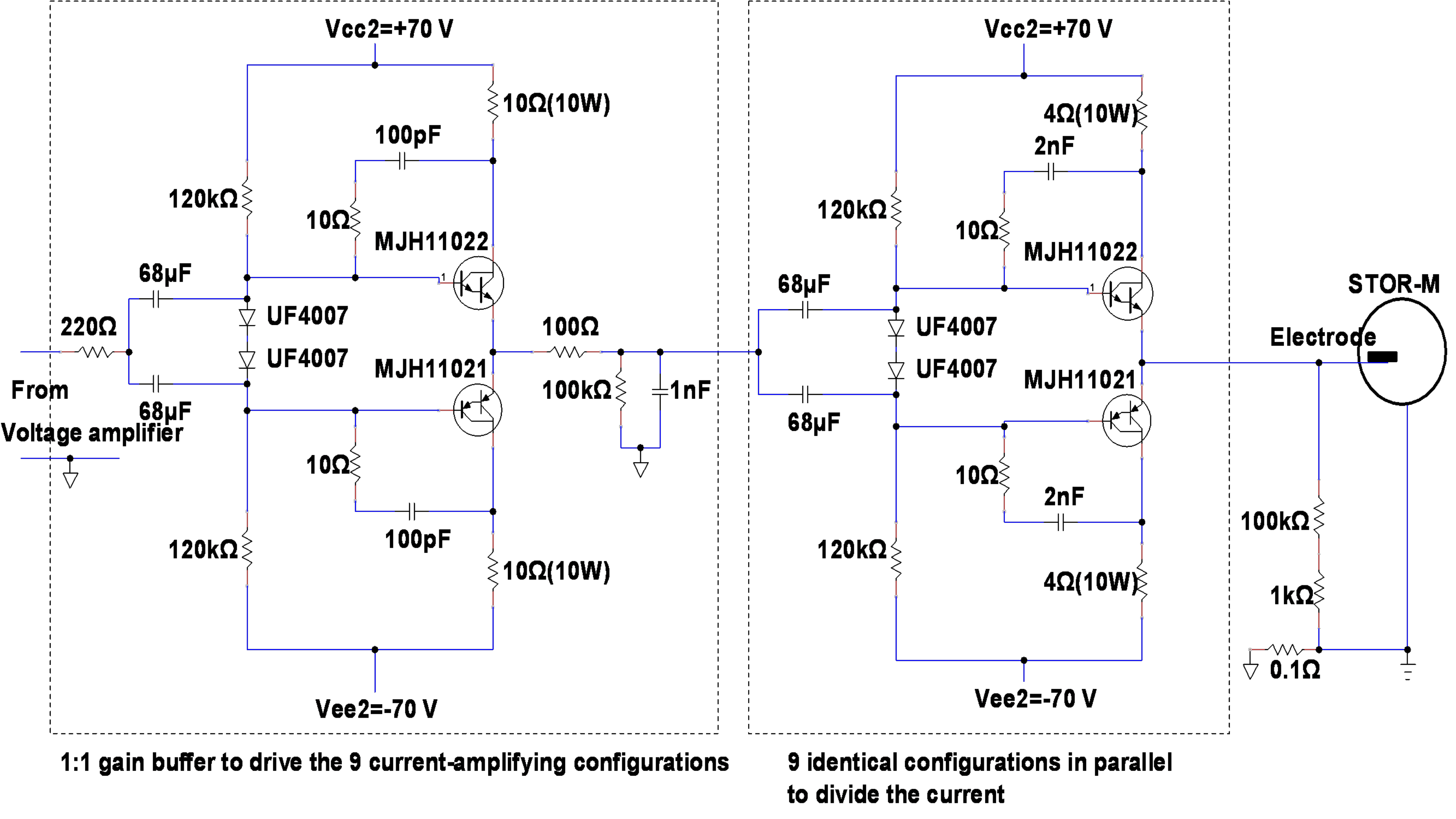}
\caption{Current amplification unit of power amplifier.} \label{fig:5}
\end{figure}
Complementary darlington silicon power transistors MJH11021 (PNP) and MJH11022(NPN) have been used for power amplification. MJH11021 \& MJH11022 have been chosen because of these two chips have maximum sustainable voltage collector-emitter$(V_{CE})$=250Vdc, collector-base$(V_{CB})$=250Vdc \& emitter-base$(V_{EB})$=5Vdc as well as maximum collector current$(I_{C})$=10Adc, base current$(I_{B})$=0.5Adc, power rating$(P_{D})$=150W, dc current gain$(h_{fe})$ $h_{femin}$=400 \& $h_{femax}$=15000, current-gain bandwidth product$(f_{T})$ $f_{Tmin}$=3MHz at ($I_{C}$=10Adc,$V_{CE}$=3Vdc,f=1MHz). Here, nine identical push-pull amplifiers has been used at the final stage for current amplification where they are connected in parallel. Circuit design had been started to control the base current of these nine identical amplifiers and maximum value of $I_{B}$ was kept within 70mA at operational condition. Here, one buffer stage has been used in between voltage amplifier and final current amplification stage to supply sufficient amount of base current. Since, the final output was planned to apply at plasma through electrode so, it is obvious that power amplifier should be capable to deliver huge amount of electron current in positive half of oscillatory signal. Here, it is designed maximum rated current 30A when DC biased voltage of each push-pull amplifier section is $\pm70$V. The typical behaviour of it is shown in figure 6. Figure 6(a) shows that typical temporal profile of input and output voltage signal. This shows
\begin{figure}[h]
\center
\includegraphics[width=260pt,height=180pt]{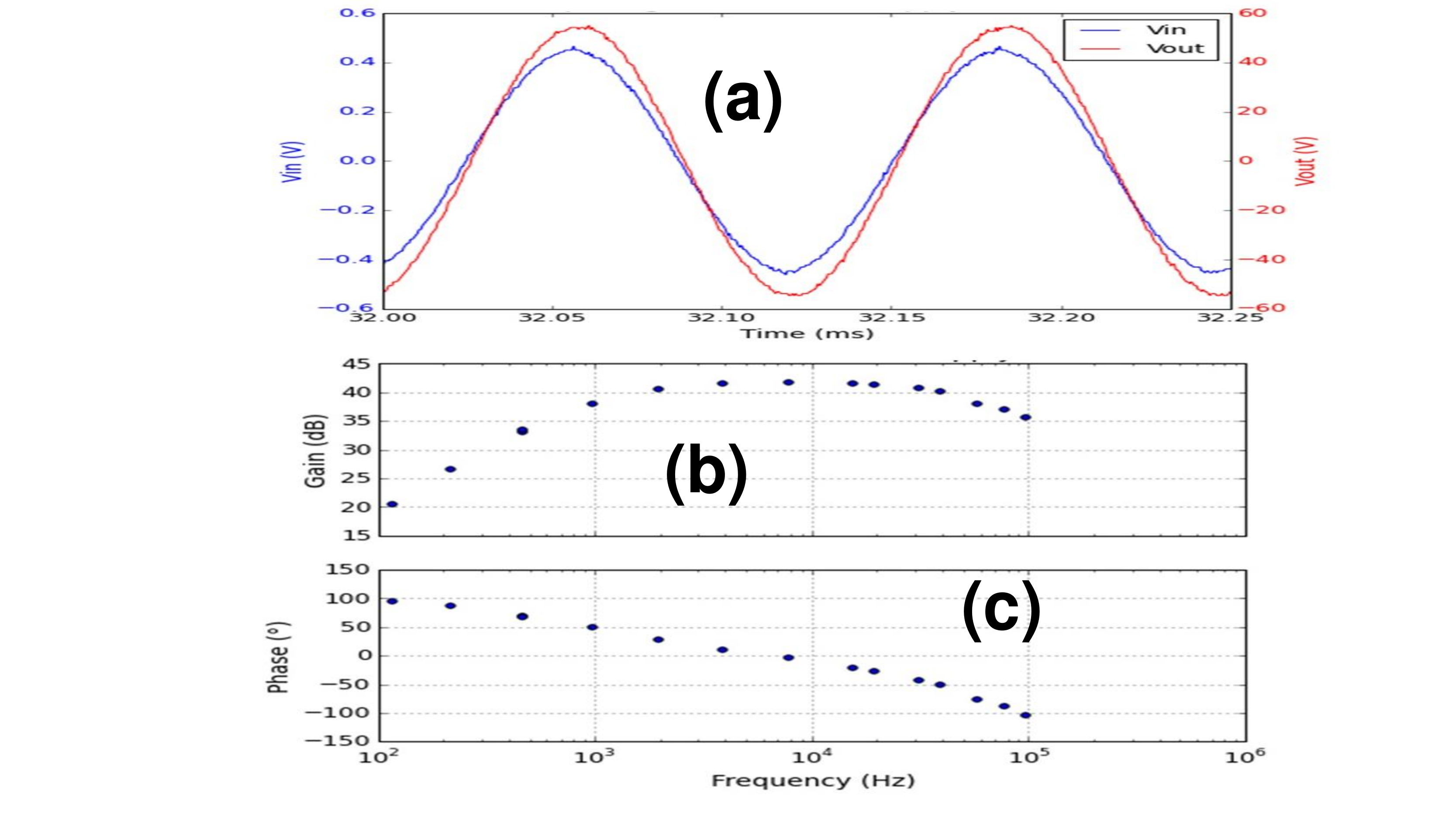}
\caption{Typical example of (a) input and output signal with time, (b)frequency versus gain, (c)frequency versus phase in between input and output signal   of power amplifier.} \label{fig:6}
\end{figure}
output signal peak value is 56V \& if $V_{CE}$=1Vdc according to characteristic of chip then calculated value of current is 3.25A since supply voltage is $\pm70$V with current limiting resistance is 4$\Omega$(10W). Therefore, total current rating of the power amplifier is $\sim$30A. Successful bench test of power amplifier in pulse mode was done through output resistance with minimum value 2$\Omega$(2000W) where no distortion of output signal has been observed. If more current is planned to draw from the circuit without any distortion in this configuration then biasing voltage of this section need to increase more than $\pm70$V according to proper calculations. Here, crossover distortion at zero voltage point was overcome by two identical diodes UF4007 with two identical biasing capacitor having capacitance 68$\mu$F. Here, current gain, calculated from its characteristic is around $h_{fe}$=1500. Figure 6(b) shows that overall frequency response of the power amplifier and it is cleared that its power gain(A) is around 43dB within frequency range 1kHz-50kHz. Here, mathematical formulation of power gain(A)=10$\log\frac{P_{out}}{P_{in}}\simeq 20\log\frac{V_{out}}{V_{in}}$, where, input resistance$(R_{in})$$\simeq$output resistance$(R_{out})$($\frac{R_{out}}{R_{in}}=2.5$). Figure 6(c) shows its phase variation in between input \& output signals which lies in between $\pm 50^{0}$ for the frequency range 1kHz-50kHz.
\section{Results \& Discussions:}
The power amplifier was successfully used for high frequency biasing experiment\cite{kn:deb1} or single frequency mode operation where improved confinement was achieved efficiently. In real time feed back experiment, signal of floating potential within selected time window has been fed into biasing electrode. The flow chart of this study has been shown in figure 1. It indicates floating potential signal from Langmuir probe is taken through an optical isolation to avoid ground loop formation which is passed through a band pass filter with lower \& upper cut off frequency 5kHz \& 40kHz respectively. Ideally this filtered frequency should passed through a broad band phase shifter but in our case it is directly connected to the input of power amplifier. Typical shot of real time feed back is shown in figure 7 where it shows that selected time window $13ms$-$21ms$ of floating potential has been fed into plasma.
\begin{figure}[h]
\center
\includegraphics[width=260pt,height=180pt]{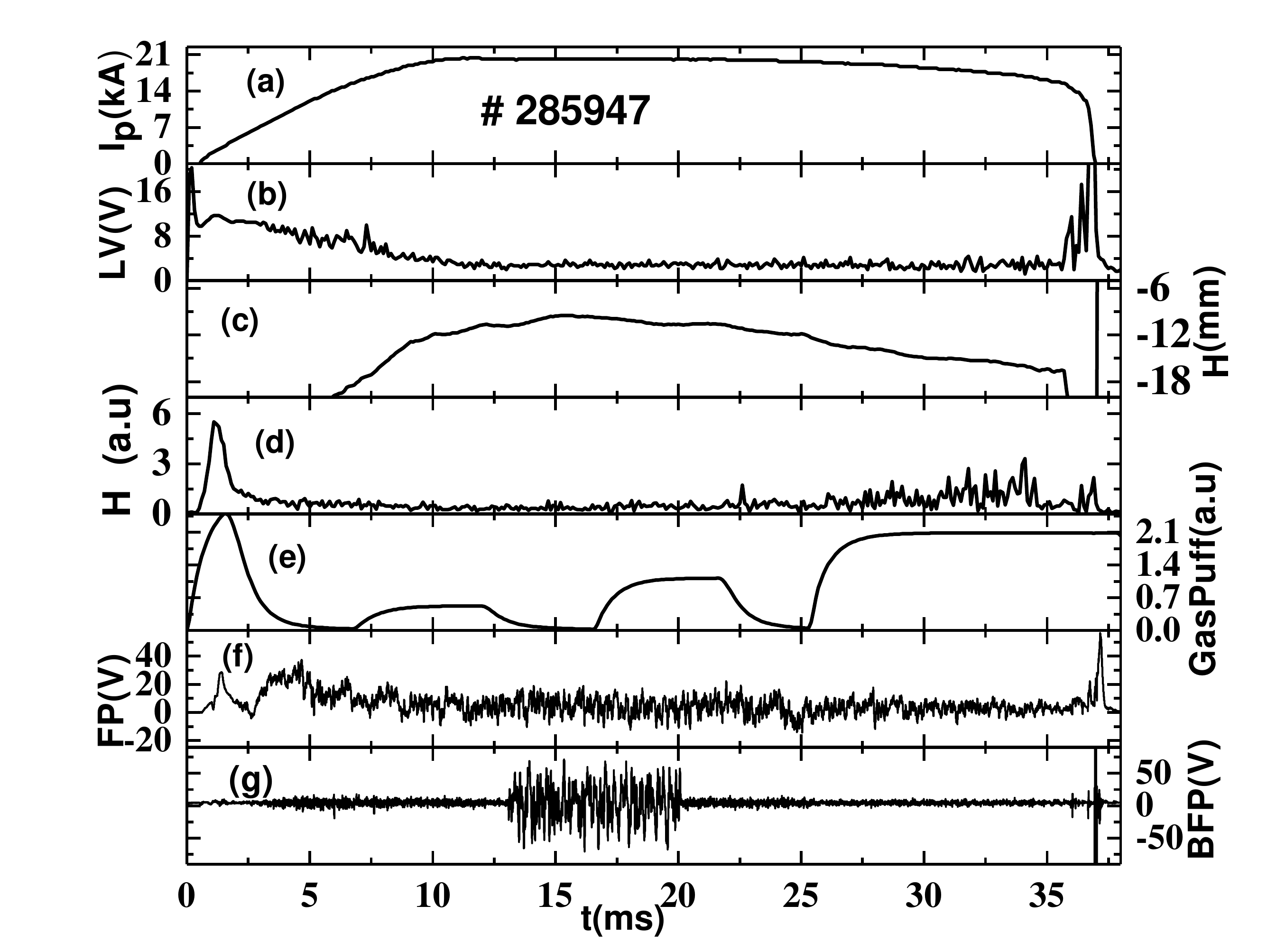}
\caption{Temporal variation of (a) plasma current, (b) loop voltage,(c)plasma horizontal position, (d)plasma $H_{\alpha}$ intensity, (e) gas puff, (f) floating potential and (g) feedback of floating signal.} \label{fig:7}
\end{figure}
$H_{\alpha}$ intensity level is not changed remarkably within active turbulence feedback time window comparing with signal level of before and after feedback. Initially, it is tried to observe mainly that whether any perturbation arises or not during application of active feedback such that plasma condition may be deteriorated. Both floating potential fluctuations as well as Mirnov signal fluctuations have been studied. Comparison is taken in between three fluctuations regimes which are before feedback, within feedback and after feedback. It is noticed that floating potential fluctuation levels within feedback time window is not changed comparing with before and after feedback fluctuations level. It is shown in figure 8.
\begin{figure}[ht]
\center
\includegraphics[width=220pt,height=180pt]{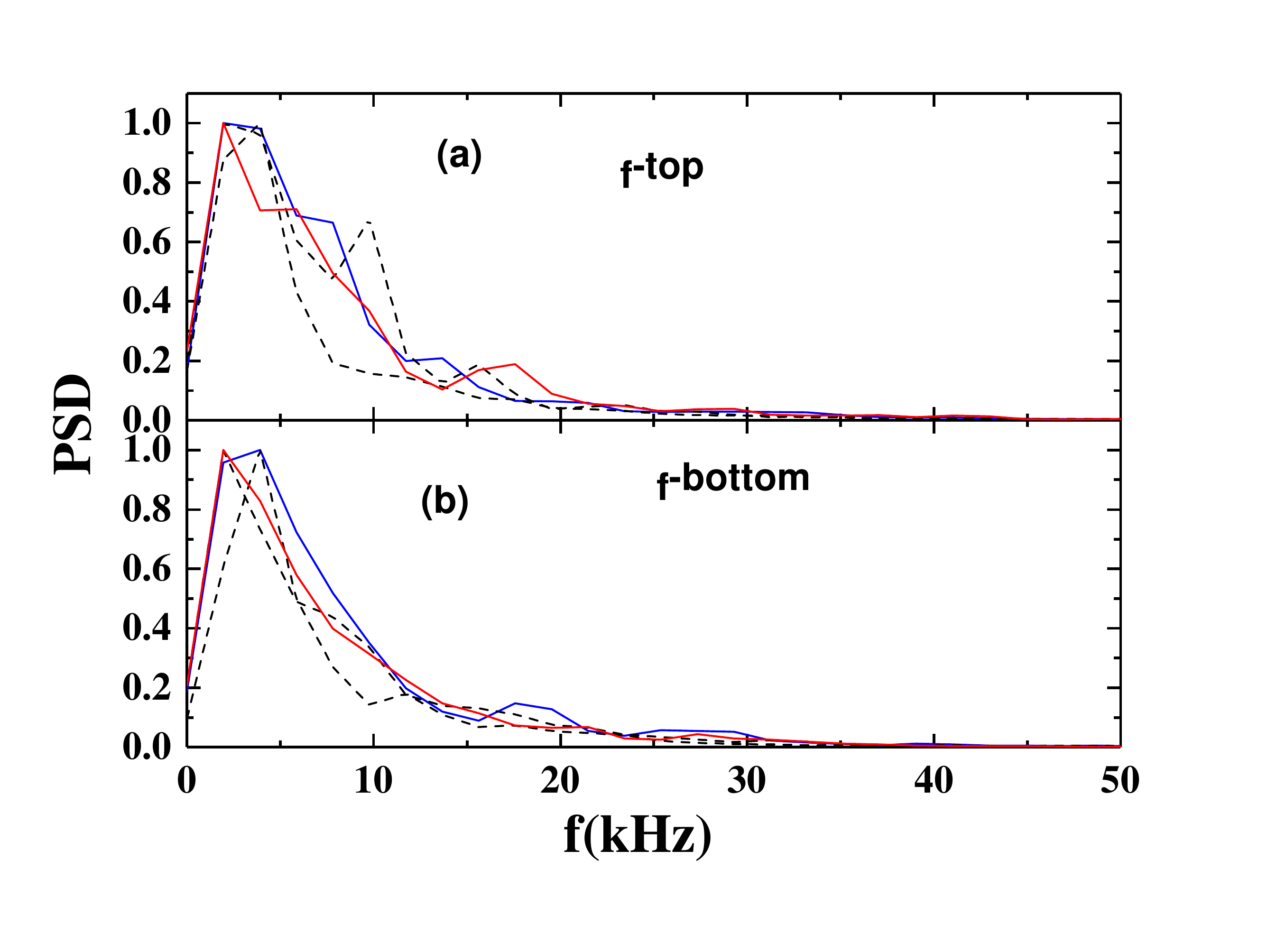}
\caption{Frequency dependance of PSD of (a) top, (b) bottom floating potential signals before(red solid), during(two black dotted) \& after(blue solid) feedback.}\label{fig:8}
\end{figure}
Here, phase of active feed back signal was drifted with respect to selected time window of floating potential which is shown in figure 9(b). Here, it is noticed that feedback signal is completely coherent within applied frequency bandwidth 5kHz-40kHz with selected floating potential time window which is fed into plasma, shown in figure 9(a) but phase is drifted from $0^{0}$ to $-90^{0}$, shown in figure 9(b).
\begin{figure}[h]
\center
\includegraphics[width=260pt,height=180pt]{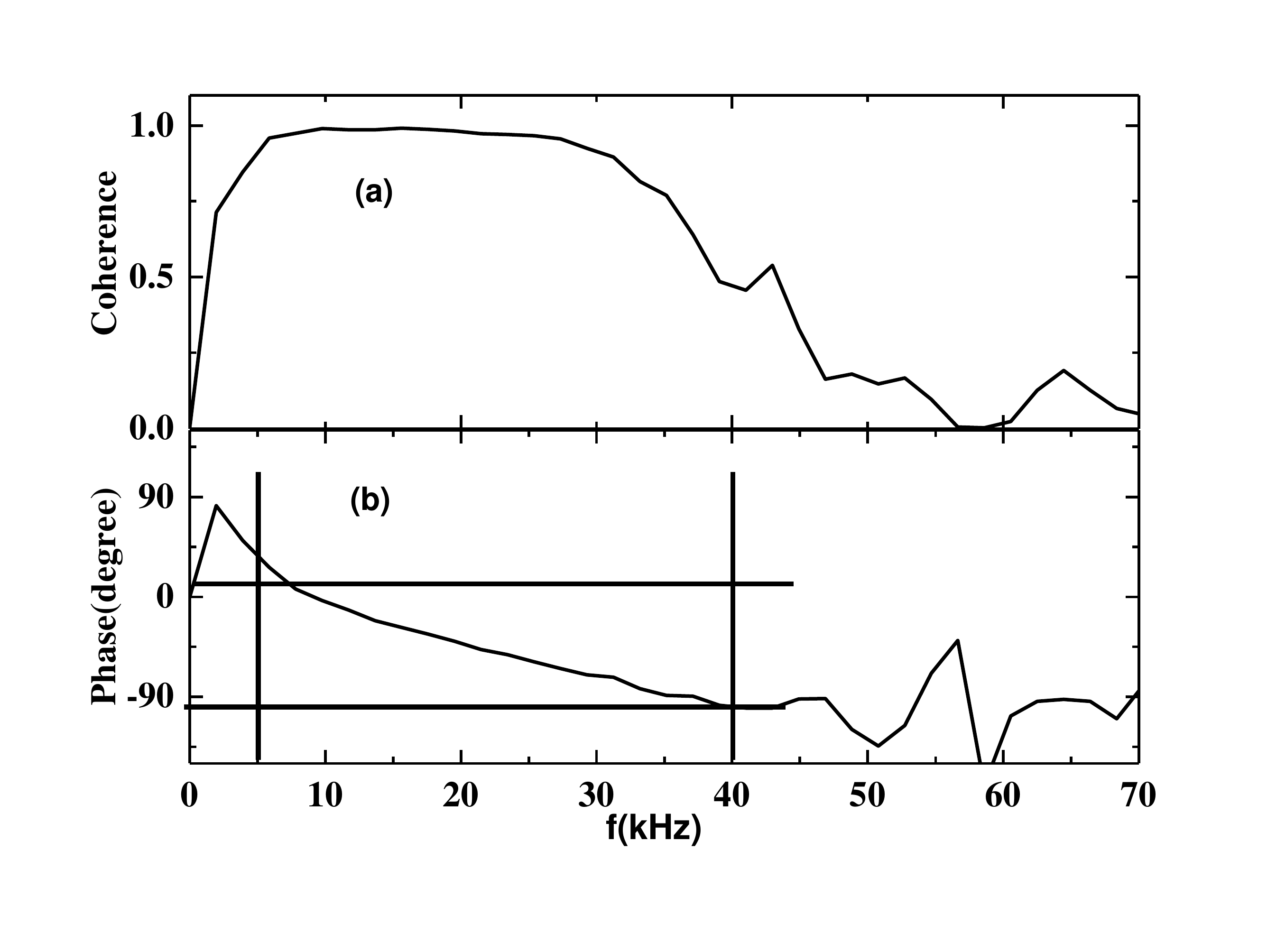}
\caption{Frequency dependance of (a)cross-coherence, (b) phase of floating potential signal and feedback signal.} \label{fig:9}
\end{figure}
\par Figure 8 clearly indicates PSD of top and bottom floating potential fluctuations before (red solid line) and after (blue solid line)feedback have no change comparing with PSD of feedback regime(dotted lines)of same signal. Here, $3ms$ time window is chosen for PSD calculations for each case. But, interestingly it is noticed that magnetic fluctuations suppression happens within feedback time window comparing same with before and after feedback at overall plasma poloidal cross-section. It is cleared from figure 10(a),10(b),10(c),10(d) that PSD of Mirnov signals from radially outboard, top, radially inboard and bottom were suppressed during feedback(red solid line) regime comparing with regime of before feed back(black solid line).
\begin{figure}[h]
\center
\includegraphics[width=260pt,height=180pt]{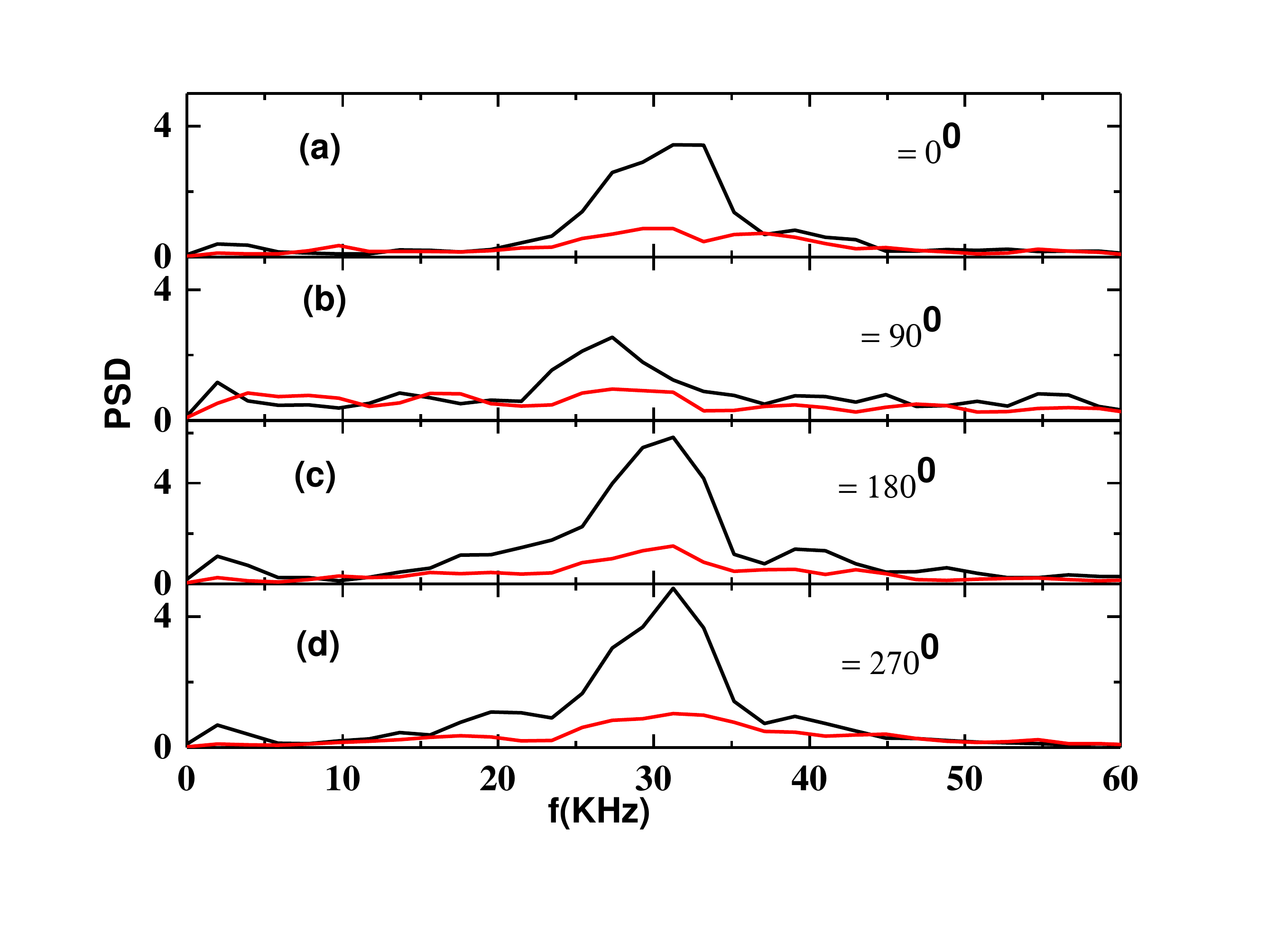}
\caption{Frequency dependance of PSD of (a)radially outboard, (b)top, (c) radially inboard, (d)bottom Mirnov signals before(black solid)\& within(red solid)feedback.} \label{fig:10}
\end{figure}
\begin{figure}[h]
\center
\includegraphics[width=260pt,height=180pt]{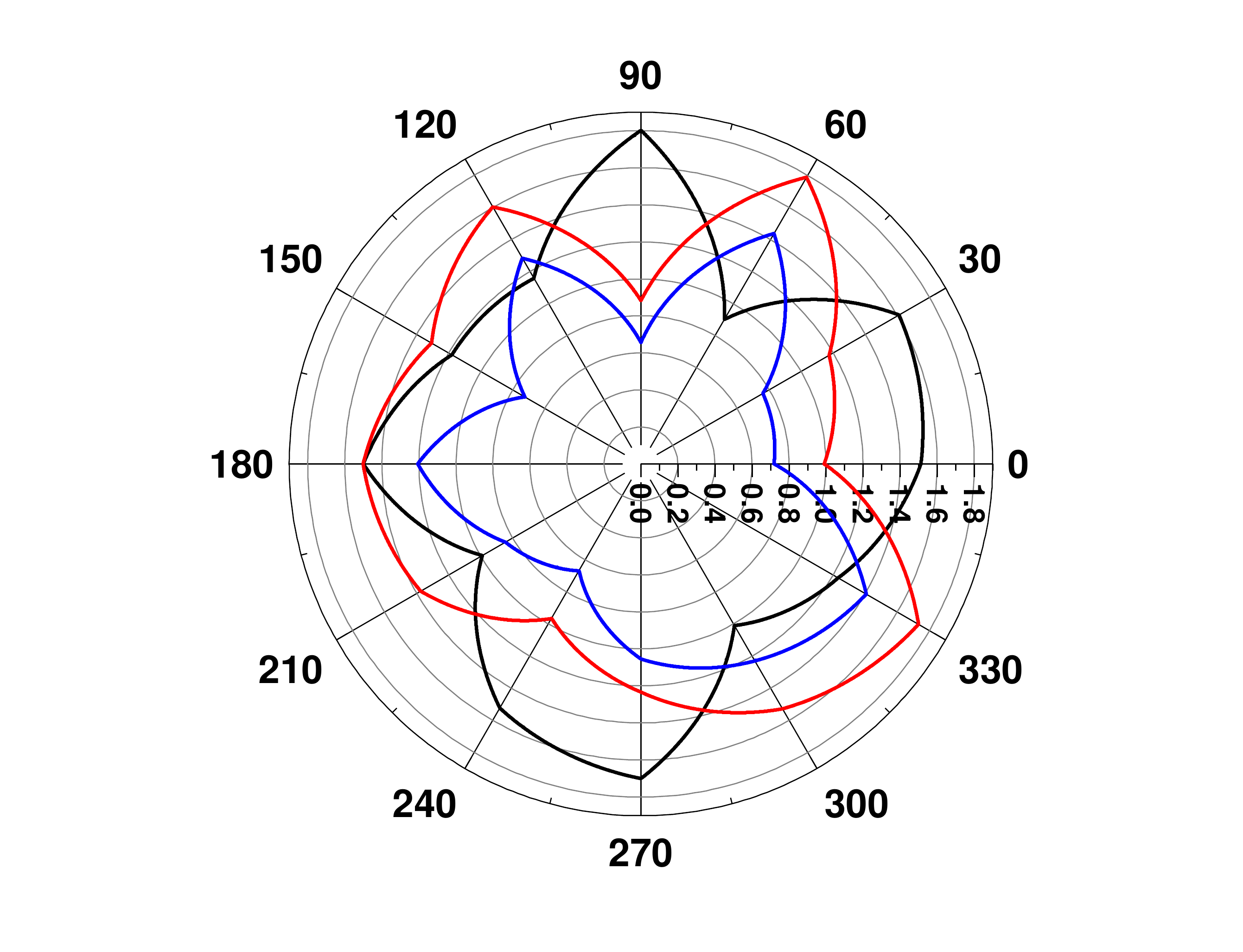}
\caption{Evolution of spatial mode structure before(black solid), during(red) \& after (blue solid)feedback.} \label{fig:11}
\end{figure}
Feedback of floating potential stabilize magnetic mode which is shown in figure 11. It shows that spatial mode structure $(m=4)$ has been stabilized. Its structure remain unchanged and it is not grown up. Mode structure evolution has been shown in figure 11 at before feedback phase(black line) $10ms-13ms$, during feedback phase(red line) $13ms-16ms$ \& after feedback phase(blue line) $17ms-20ms$.
\section{Conclusion:}
Successful active feed back turbulent experiment has been performed in STOR-M tokamak with free phase drifting. Interestingly, it is noticed that plasma electrostatic turbulence is not enhanced during feedback time window but magnetic fluctuations has been suppressed due to active feedback. In future, phase shifter will be used to active control over phase to observe its effect for extensive experimental studies.
\begin{acknowledgments}
 We would like to acknowledge NSERC for supporting this work. We also would like to acknowledge machine workshop. Specially acknowledge thanks go to Mr. Chomyshen and Mr. Toporowski in the machine workshop for their kind help and friendly approach when needed.
\end{acknowledgments}


\section{References:}
\begin{thebibliography}{150}
\bibitem{kn:Brave}R. Y. Bravenec, et.al., Nucl.Fusion \textbf{31}, 687(1991)
\bibitem{kn:Arsenin}y. Y. Arsenin, et.al., JETP Lett. \textbf{8}, 41{1968}
\bibitem{kn:Parker}R. R. Parker and K. I. Thomassen, Phys. Rev. Lett. \textbf{22}, 1171{1969}
\bibitem{kn:Keen}B. E. Keen and R. Y. Aldridge, Phys. Rev. Lett. \textbf{22}, 1358{1969}
\bibitem{kn:Bol}K. Bol, J. L. Cecchi, et. al., plasma Physics and Controlled Nuclear Fusion Research (Proceedings of the Fifth International Conference, Tokyo) \textbf{1}, 83(1974)
\bibitem{kn:Liewer1}P. C. Liewer, Nucl. Fusion  \textbf{25} 543(1985)
\bibitem{kn:Callen}J. D. Callen, Phys. Fluids B \textbf{4} 2142 (1992)
\bibitem{kn:UCKAN}T. UCKAN et al., Nucl. Fusion \textbf{35}, 487 (1995)
\bibitem{kn:Kan}Zhai Kan et al., Physical Review E \textbf{55} 3431 (1997)
\bibitem{kn:Brooks}J. W. Brooks, et.al., Review of Scientific Instruments \textbf{90}, 023503 (2019)
\bibitem{kn:deb1}Debjyoti Basu, et.al., ``Application of high frequency biasing and its effect in STOR-M tokamak'', manuscript submitted in Nuclear Fusion
\bibitem{kn:deb2}Debjyoti Basu, et.al., Nucl. Fusion \textbf{58}, 024001 (2018)
\bibitem{kn:Xiao1}C. Xiao, et.al., Review of Scientific Instruments \textbf{79}, 10E926 (2008)
\end{thebibliography}
\end{document}